\documentclass[sigconf]{acmart}

\renewcommand\footnotetextcopyrightpermission[1]{} % removes footnote with conference information in first column
\usepackage{textcomp}
\usepackage{xcolor}
\usepackage{xspace}
\usepackage{caption}
\usepackage{multirow}
\usepackage{algorithm}
\usepackage{algorithmicx}
\usepackage[noend]{algpseudocode}
\usepackage{booktabs} % For formal tables

\def\BibTeX{{\rm B\kern-.05em{\sc i\kern-.025em b}\kern-.08em
    T\kern-.1667em\lower.7ex\hbox{E}\kern-.125emX}}

\newcommand{\para}[1]{\xspace \smallskip \noindent\textbf{#1}\xspace}

\author{Yikun Ban}
\affiliation{%
  \institution{Peking University \& University of Illinois at Urbana-Champaign}
}
\email{yikunb2@illinois.edu}

\author{Yuchen Zhou}
\affiliation{
	\institution{Peking University}
}
\email{yuchen_bubble@pku.edu.cn}

\author{Xu Cheng}
\affiliation{
	\institution{Peking University}
}
\email{chengxu@mprc.pku.edu.cn}

\author{Jiangfang Yi}
\affiliation{
	\institution{Peking University}
}
\email{yijiangfang@pku.edu.cn}

\title{Coalesced TLB to Exploit Diverse Contiguity of Memory Mapping}

\begin{document}

%%should not be used}

\thispagestyle{empty}

\begin{abstract}
\vspace{0.3cm}
The miss rate of TLB is crucial to the performance of address translation for virtual memory. To reduce the TLB misses, improving translation coverage of TLB has been an primary approach. Many previous works focus on coalescing multiple contiguously mapped pages of the memory mapping into a modified entry, which function well if the assumed contiguity of memory mapping is given. Unfortunately, scenarios of applications are complicated and the produced contiguity diversify. To gain better performance of translation, in this paper, we first introduce a complex but prevalent type of contiguity, mixed contiguity. Then we propose a HW-SW  hybrid coalesced TLB structure which works well on all observed types of contiguity including this type. In our evaluation, the proposed scheme, \textbf{K}-bit Aligned TLB, outperforms the state-of-the-art work by reducing at lease 27\% TLB misses on average over it  using 16 benchmarks.

\end{abstract}

\maketitle

\section{Introduction}

Virtual memory is universally used  for CPUs  and increasingly for the accelerators like GPUs to date\cite{CPU,gpu1,DVM}. 
%It constructs an illusion of private and adequate address space for each process, improving the programming productivity, memory  protection, and communications of cores. 
Translation Lookaside Buffer (TLB) is vital to the translation performance from virtual to physical address.
Unfortunately, with the escalation of memory capacity demands for large memory applications, the overheads of address translation have worsened since the coverage of translation of TLB become more limited.  For many big-memory workloads, the overheads of address translation can reach up to 50 \% of execution time \cite{efficient,bhattacharjee2013large,karakostas2014performance}.
To reduce the cost caused by page table walk, many research foci have intensified on improving translation coverage of TLB \cite{efficient,THP2,NVM2,RMM,predict1,Efficientsynonym,cluster,colt,zhang2010enigma,anchortlb,DVM},  motivated by exploiting the \textit{contiguity} existed in memory mapping.

We refer to contiguity as describing the distribution of contiguity chunks among memory mappings for a process and a contiguity chunk denotes a range of pages in which both virtual and physical addresses are contiguously mapped (see Definition \ref{def1}). In fact, as the sizes of contiguity chunk allocated by Operating System(OS) are disparate, the contiguity may be divided into several types. The first common approaches exploit large contiguity utilizing huge pages \cite{huge1,huge2,huge3}. For example, the common x86 architecture supports using 2MB page size to represent hundreds of base 4KB pages. However, these works only cover large chunks with discrete sizes and thus the scalability of coverage improvement is insufficient. Direct segment \cite{efficient} and RMM\cite{RMM} start adding variable-sized HW segments to displace page-based translations. For small contiguity, HW coalescing techniques\cite{cluster,colt} are proposed to compress up to 8 contiguous pages into an TLB entry, often being used for fragmented memory allocations.

It has been examined that each coalescing approach mentioned above can not achieve the expected performance other than being provided with suitable type of contiguity (see Section \ref{sec:mix} ). When encountering unsuitable contiguity chunks, be it too small or too large, the prior coalescing mechanisms are no longer efficient. For example, a 256 $\times$ 4KB (1MB) contiguity chunk is too small to be placed into a huge page (2MB), and too large to be coalesced into an entry by HW coalescing methods. Afterwards, Anchor\cite{anchortlb} proposed the hybrid coalescing technique augmenting the intervention of OS, in order to dynamically change the optimal fitting contiguity with the assumption that there only is one type of contiguity in memory mapping (i.e., the sizes of contiguity chunks are highly similar). Unfortunately, this assumption does rarely exist. We have examined tens of widely used benchmarks such as Spec 2006 suites\cite{spec2006}, Graph 500\cite{efficient}, Parsec\cite{parsec}, and BioBench suites\cite{biobench}, and find that the memory mappings of more than 90\% workloads contain more than one type of contiguity, which we call \textit{mixed contiguity}. The observation indicates that Anchor's performance is constrained by the mixed contiguity in many applications with same reason as prior coalescing approaches.

In this paper, we propose a HW-SW hybrid TLB coalescing structure, $\mathbf{K}$-bit Aligned TLB,  beyond Anchor \cite{anchortlb} by exploiting different types of contiguity in a memory mapping simultaneously, further improving the translation coverage of TLB.
While OS is allocating memory, this work utilizes multiple types of aligned page table entries, denoted by $\mathbf{K}$, to cover different sizes of contiguity chunks. In detail, a type of aligned entries (e.g., $k$-bit aligned) represent a subset of page table entries where the LSB $k$ bits of the Virtual Page Number (VPN) are clearing out, which store the local contiguity by recording the number of contiguous pages in the next $2^k$ pages. After the page table walk, in addition to return the requested Physical Page Number (PPN) to CPU, OS will select the optimal matching aligned entry with the maximal coverage to fill up L2 TLB if contiguity matches. Then, after L1 TLB misses, if the requested VPN is not in the L2 TLB, its corresponding aligned entry is used to translate VPN by simply adding the virtual address difference between the aligned and requested pages, to the physical page address of the aligned entry.

In the hybrid coalescing, L2 look-up is divided into two steps: \textit{regular look-up} and \textit{aligned look-up}.
As support filling multiple types of aligned entries, the aligned look-up that searches the matching aligned entry becomes complicated and may be relatively time-consuming.  To reduce the overheads of aligned look-up, this work leverage spatial locality of memory accesses to predict which type of alignment the requested VPN belongs to, significantly reducing the overheads of aligned look-up. Moreover, this work proposes a heuristic algorithm to determine $\mathbf{K}$ at the initialization of memory allocation for a process, adapting to diverse memory allocation scenarios. Empirically, $\mathbf{K}$ is robust to tolerate the dynamic change of memory mappings and maintain efficiency meanwhile, minimizing the cost of updating page table\cite{anchortlb}. 

In summary, the main contributions of this paper are:
   
\begin{itemize}
\item We demonstrate the prevalence of mixed contiguity in diverse workloads and discuss the limitations of prior TLB coalescing techniques for this type of contiguity.
    
\item We propose K-bit Aligned TLB, a hybrid coalescing scheme, which adaptively assigns each contiguity chunk a optimal container for coalescing, maintaining efficiency in face of diverse types of contiguity including mixed contiguity. 

\item We use real-machine traces to evaluate all approaches on a broad range of workloads. The proposed scheme decrease the TLB misses by 36 \% on average over the state-of-the-art approach Anchor \cite{anchortlb}.

\end{itemize}

\begin{figure}
\centering
 \includegraphics[width=1.0\columnwidth]{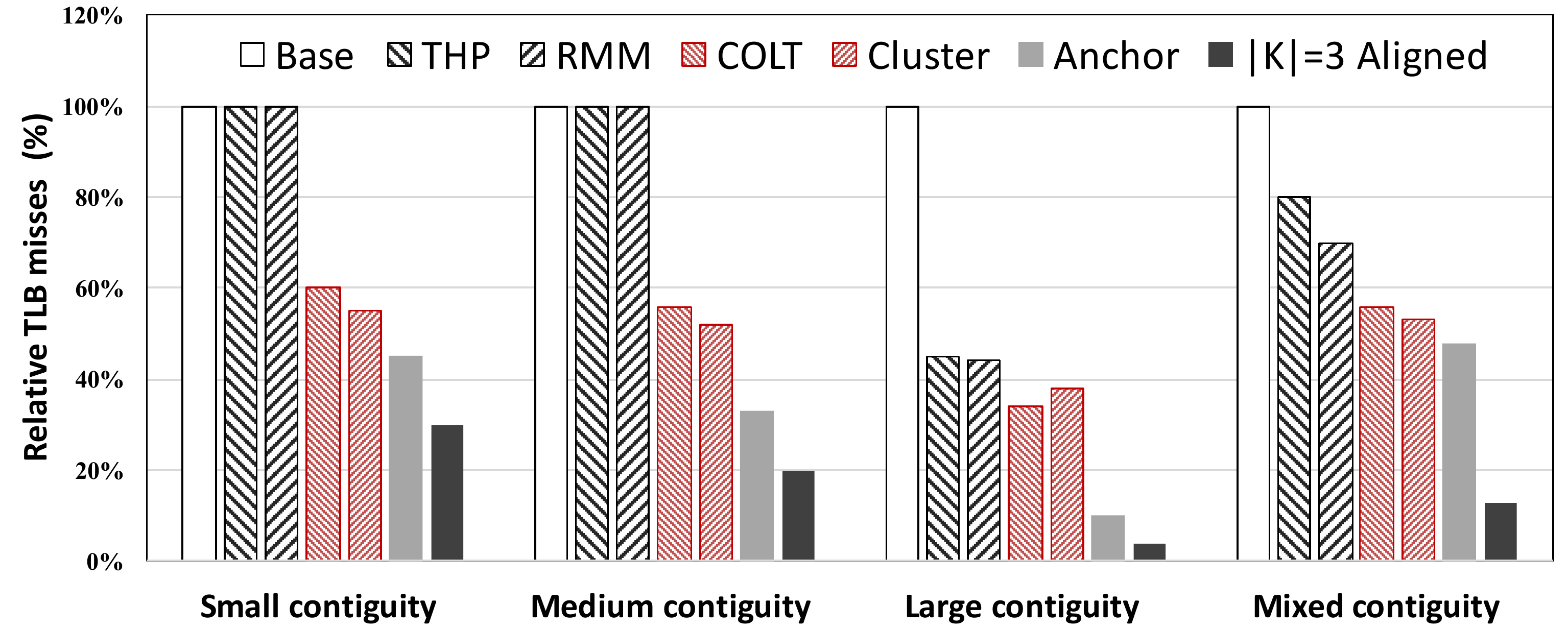}
  \vspace{-0.2cm}
 \caption{Relative TLB misses of existing techniques for four different types of contiguity} 
 \label{fig:misscon}
% \vspace{-0.4cm}
\end{figure}

\begin{figure*}[t]
\centering
 \includegraphics[width=1.0\textwidth]{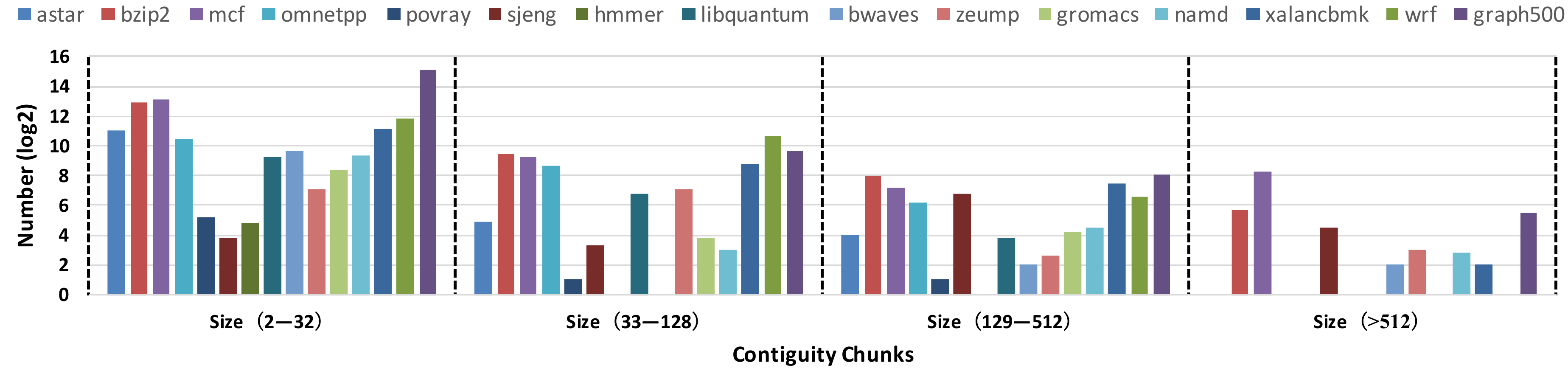}
  \vspace{-0.3cm}
 \caption{THP\cite{THP} off. The distribution of contiguity chunks at first billionth instruction boundary for the used benchmarks. Y-axis is the number of contiguity chunks (denoted by $n$) displayed by $log_2^{n+1}$; X-axis is the four types of contiguity chunk classified by the size. 14 out of 15 benchmarks have more than one type of contiguity, which all can be thought of the mixed contiguity. }
 \label{fig:cc}
 %\vspace{-0.4cm}
\end{figure*}

\begin{figure*}[t]
\centering
 \includegraphics[width=1.0\textwidth]{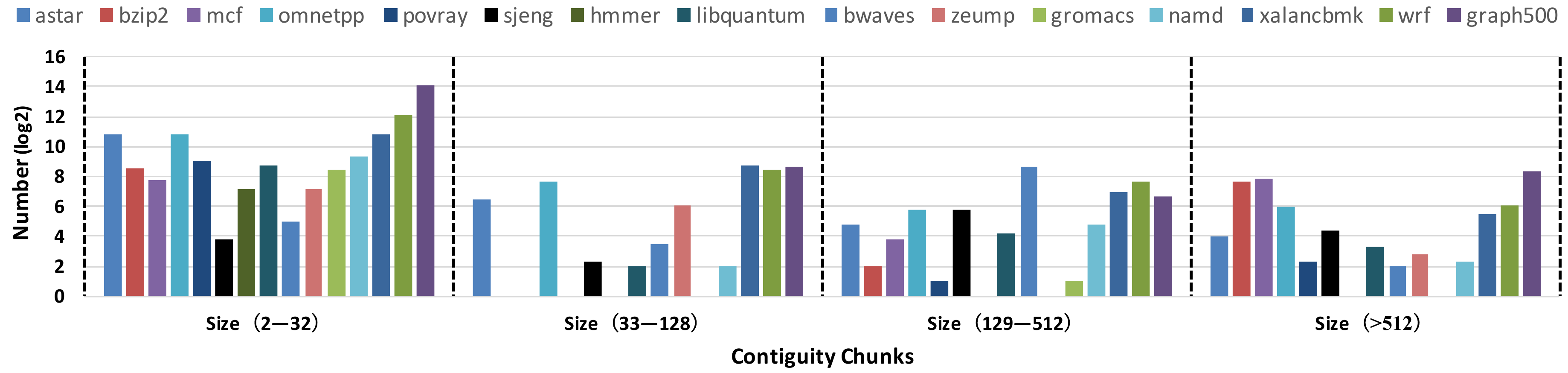}
  \vspace{-0.3cm}
 \caption{THP\cite{THP} on. The distribution of contiguity chunks at first billionth instruction boundary for the used benchmarks. Although enable OS to support huge pages, 14 out of 15 benchmarks still have the mixed contiguity. }
 \label{fig:cc2}
 %\vspace{-0.4cm}
\end{figure*}

\section{Contiguity in Memory Mapping}\label{sec:back}

In this section, we first analyze the contiguity and then elaborate an existing challenge. 
Due to the buddy allocation mechanism of operating system,  the memory mappings for the plenty of applications exhibit the contiguity in different degrees\cite{colt}. With this diversity of behaviors, many works were proposed to leverage the corresponding form of contiguity, improving the translation coverage of TLB. 
To better deliver our points, we use the definition of contiguity chunks.
\begin{definition} \label{def1} (Contiguity Chunk and Size)
A contiguity chunk denotes a number of pages in page table in which their both virtual and physical addresses are contiguously mapped, and it does not exist that a contiguity chunk is contained by another contiguity chunk. The size of a contiguity chunk is the number of pages denoted by the chunk.
\end{definition}

In figure \ref{fig:ptxx}, for example,  three contiguity chunks occur in the page table and their sizes are 2, 3 and 6.

\subsection{Diversity of Contiguity}

To begin with, the existing contiguity of memory mappings may be divided into four types. The first type of contiguity is \textit{large contiguity} composed of contiguity chunks of size more than hundreds.  Huge pages\cite{huge1,huge2,huge3} focus on this type of contiguity, relying on the allocation of large contiguity chunks by OS. For instance, the x86-64 architecture supports using 2MB and 1GB huge pages to substitute 512  and 262,144 contiguously mapped 4KB pages respectively. 
In spite of its effectiveness, huge pages usually cannot reach the optimal coverage improvement produced by a contiguity chunk.  
Considering a contiguity chunk of size between 512 and 1024, only a portion of contiguity is able to be coalesced because of the discrete huge page sizes.    
Direct segment\cite{efficient} and RMM\cite{RMM} introduce segments to completely cover contiguity chunks, precluding the page size limitation. However, as the HW of segment TLB is fully associative and only has a limited number of entries, each segment is expected to represent a very large contiguity chunk and thus OS is required to make non-trivial changes for maintaining this invasive allocation.

Opposite to large contiguity, the second type of contiguity is \textit{small contiguity}, which consists of contiguity chunks with tens of pages. This type of contiguity is prevalent among fragmented memory mappings. In long-running system, large contiguous regions of memory are often fragmented to small and varying size of contiguous regions, because the in-use pages distributed among memory inhibit the allocation of large contiguity chunks \cite{rangar}. Moreover, the common NUMA architectures, such as 3D stacked DRAMs\cite{3dstack}, network-connected hybrid memory cube (HMC)\cite{hmc},  and non-volatile memory (NVM)\cite{dulloor2016data, NVM2,NVM3}  require fine-grained memory mapping to place frequently accessed pages on fast near memory, rising the non-uniformity memory and aggravating the fragmentization of memory\cite{anchortlb}.  The HW coalescing techniques: Clot\cite{colt} and Cluster\cite{cluster} are proposed to account for the small contiguity. They fetch up to 8 page table entries by a cache line and try to coalesce them into one entry. Naturally, as the size of contiguity chunk increases, the number of coalesced entries to cover this chunk mounts. With the restriction of TLB HW, the scalability of their coverage will be limited as the augment of contiguity.

The third type of contiguity is the medium of large and small contiguity, called medium contiguity. This type of contiguity describes the distributed contiguity chunks which falls short of huge pages while is larger than fine-grained memory mappings. Previous works have observed the medium contiguity on many real-world applications \cite{colt,cluster,anchortlb}.

As the variance of applications and system status, \cite{anchortlb} proposed the hybrid TLB coalescing to adapt to the these types of contiguity respectively. It introduces the anchored page table: use anchor entries (can be thought of a type of aligned entry) uniformly distributed among page table entries to record local contiguity and fill L2 TLB with the aligned entry instead of requested entry if contiguity matches. OS is modified to tune the density of anchor entry based on the memory mapping status. However, the cost of the adaptation is non-trivial and the eventually coverage improvement  is often limited, as the analysis of following section.

\subsection{Mixed Contiguity}\label{sec:mix}

To obtain the ideal performance, each existing TLB coalescing approach mentioned above has the suited type of contiguity. Even though for Anchor TLB \cite{anchortlb} which dynamically change the anchor distance every billion instructions to fit the current memory mapping status, it has an optimal type of contiguity asynchronously. With this fact, the performance of the existing TLB coalescing methods will degrade in face of the memory mapping containing more than one type of contiguity, \textit{mixed contiguity}. In practice, the pure memory mapping, consisting of only one type of contiguity, rarely exists in system, as the result of the disordered and fragmented memory allocation.

To quantify the variance of contiguity in a process, we periodically record the contiguity chunks among the memory mappings on a X86-64 machine with linux 4.16 after warming up. We scan the page table for a process every 1 minute using the pagemap\cite{pagemap} and running benchmarks Spec 2006.

Figure \ref{fig:cc} and \ref{fig:cc2} show the histogram of the size distribution of contiguity chunks for fifteen benchmarks. The x-axis is the average number of contiguity chunks for ten scanning, and the y-axis is the size range of contiguity chunks to represent different types of contiguity. The histogram demonstrates that the memory mapping for a process often include different types of contiguity simultaneously (i.e., contiguity chunks with various sizes).

 We have examined the limitation of existing methods for coalescing contiguity chunks with assorted sizes in a memory mapping status, as shown in Figure \ref{fig:misscon}. The method of synthesizing contiguity types is described in Section 4.1. First, Transparent Huge Page (THP) \cite{huge1}, direct segments and RMM \cite{efficient, RMM} are capable of covering large contiguity chunks (size > 512), while neglect the small and medium ones. Therefore, encountering the mixed contiguity, the ceiling of TLB coverage improvement is considerably lowered for this class of methods.Second, on the ground that one coalesced entry of HW coalescing techniques\cite{cluster,colt} represents up to 8 pages, a contiguity chunk with considerable size (e.g., 256) needs plenty of ( 32 at least) coalesced entries to be covered. Unsurprisingly, with mixed contiguity, the performance of the HW coalescing techniques is far-away from our expectations. Third, hybrid TLB coalescing \cite{anchortlb} tune the optimal anchor distance among the page table entries to fit the size of contiguity chunk. Unfortunately, it only can take care of one type of contiguity in a period. In the page table, every N (anchor distance) entries is placed an anchored entry which records the number of following contiguous pages ending to next anchored entry. For example, if memory pages are allocated in contiguity chunk of size 16, the optimal anchor distance is 16. However, referring to the contiguity chunks larger than anchor distance, multiple anchored entries are needed to be covered. With respect to the contiguity chunk smaller than anchor distance, it will be neglected if the discontinuous pages exist between the chunk and the corresponding anchored entry.

 To sum up, mixd contiguity is a usual type of contiguity distribution among memory mappings. With this contiguity, previous approaches improve the TLB coverage in varying degrees compared to conventional TLB, while is far away from our expected improvements. To address this challenge, we propose the following scheme.

\section{Method}

\begin{figure}
\centering
 \includegraphics[width=1.0\columnwidth]{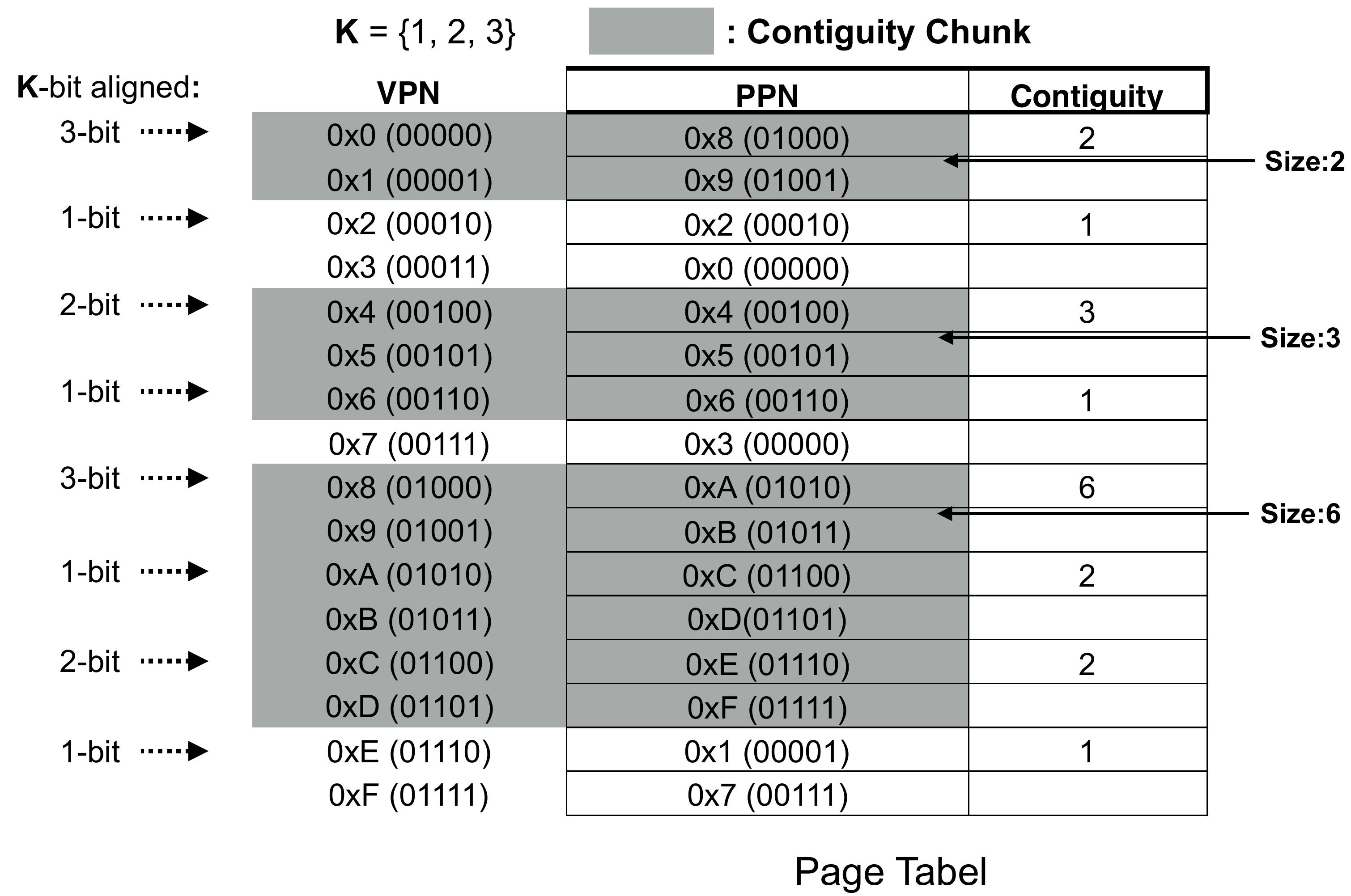}
 % \vspace{-0.2cm}
 \caption{Page Table with $\mathbf{K}$-bit Alignments} 
 \label{fig:ptxx}
 %\vspace{-0.4cm}
\end{figure}

To maximize the TLB coverage in face of the skewed distribution of contiguity in memory mapping, in this section, we propose  an effective solution for improving the TLB coverage with a few added bits for each TLB entry and a slight modification of operating system.

\subsection{Page Table with $\mathbf{K}$-bit Alignments }

To coalesce a bunch of contiguity pages into one entry, each page table entry is assigned a property value, \textit{contiguity}, representing the number of following pages (including the entry itself) in which both VPNs and PPNs are contiguous. The contiguity is stored in the unused bits of the page table entry, as shown in Figure \ref{fig:entry}. 

%too complex to understand
In our approach, instead of designating uniformly distributed entries to record how many pages are contiguously mapped\cite{anchortlb}, we use multi-granular distributed ($\mathbf{K}$-bit aligned) entries in the entire page table for more complicated memory mappings. 

Given a natural number $k$ that denotes a type of alignment,  we use $k$-bit aligned entries to represent all the entries among the page table entries in which the $k$ LSB bits of the VPN are zero. If an entry is $k$-bit aligned, in addition to store conventional information, it is designated as recording the number of pages that are contiguously mapped for the following $2^k$ pages, starting from the aligned entry. As more types of alignment is added to page table (e.g., $k$-bit, $h$-bit),  $\mathbf{K}$ is a set of natural numbers (e.g., $\mathbf{K} = \{k, h\}$) to denote all the types of alignments placed in the page table.   

Figure \ref{fig:ptxx} shows a page table and its $\mathbf{K}$-bit aligned entries containing 1-bit, 2-bit and 3-bit aligned entires. In the example, VPN 4 is located at a 2-bit aligned entry and the recorded contiguity is 3 which means three pages including itself are contiguously mapped in the following 3 pages; VPN 8 is 3-bit aligned and contiguity of the corresponding entry is 6, completely covering the contiguity chunks of size 6.  An entry is easy to be decided  whether it is an aligned entry and which types of alignment it belongs to by checking the LSB bits of its VPN. Notice that VPN 8 actually is 1-bit, 2-bit or 3-bit aligned while it is defined as 3-bit aligned because $\mathbf{K}$-bit aligned entries comply with the following compatible rule.

\para{Rightward Compatible Rule.}
Let VPN$_1$ be $a$-bit aligned, VPN$_2$ be $b$-bit aligned. If $a>b$, then VPN$_1$ is also $b$-bit aligned because the LSB $b$ bit of VPN$_1$ must be zero. For example, VPN $4$ is $2$-bit aligned, and also is $1$-bit aligned.
To best capture the local contiguity information, we set an entry as a $k$-bit aligned PTE as the following rule:

\smallskip
\textit{  Given a set $\mathbf{K} = \{a, b, c ...\}$, if the VPN of an entry satisfies $\mathcal{K}$-bit aligned where $\mathcal{K} \subseteq \mathbf{K}$, we set the entry as a $\hat{k}$-bit aligned entry where $\hat{k}$ is the maximum of $\mathcal{K}$.}
\smallskip

  For example, Given a set $\mathbf{K} = \{1, 2, 3\}$ and a VPN $8$, we define this VPN as $3$-bit aligned based on Rightward Compatible Rule, in spite of the fact that VPN $8$ is ${1, 2, 3}$-bit aligned ($1$-bit, $2$-bit and $3$-bit aligned). Analogically, VPN $6$ is $1$-bit aligned; VPN $4$ is $2$-bit aligned.

%we use $\mathbf{K}$-bit Aligned PTs to store diverse contiguity chunks, where $\mathbf{K}$ is a set of some specific natural numbers determined by Algorithm $1$ in Section 5. 
%Given a natural number $k \in \mathbf{K}$, we call an entry as $k$-bit Aligned PTE when the right $k$ bits of its VPN all are \emph{zero}. 
%If an entry is $k$-bit aligned, in addition to store conventional information, it is designated as recording the number of following pages that are contiguously mapped. 
%Note that the maximal number of contiguous pages it can record is $2^k$.

In fact, one type of aligned entry is only able to optimally cover one type of contiguity chunk size. 
For instance, assuming a contiguity chunk of size  $32$, its contiguity information can be completely stored in a $5$-bit aligned entry. 
However, it needs to consume at least eight $2$-bit aligned entries for achieving this status. If we choose a bigger bits aligned entry, like $7$-bit aligned, it is prone to lose some other contiguity chunks in the address range covered by this entry. 
To overcome this problem, therefore, we propose $\mathbf{K}$-bit Aligned PTEs, leverage multiple types of aligned entries to encode the information of diverse sizes of contiguity chunks.

\subsection{Translation using $\mathbf{K}$-bit Aligned Entries}
As an aligned entry represents a range of contiguous pages, holding aligned entries in the TLB can boost the translation coverage. Besides regular entries, to translate with $\mathbf{K}$-bit aligned entries, a few additional bits are added to per entry of TLB. Considering the sensitivity of L1 TLB to access latency, we carry out this design on L2 TLB.

\begin{figure}[ht]
\centering
 \includegraphics[width=1.0\columnwidth]{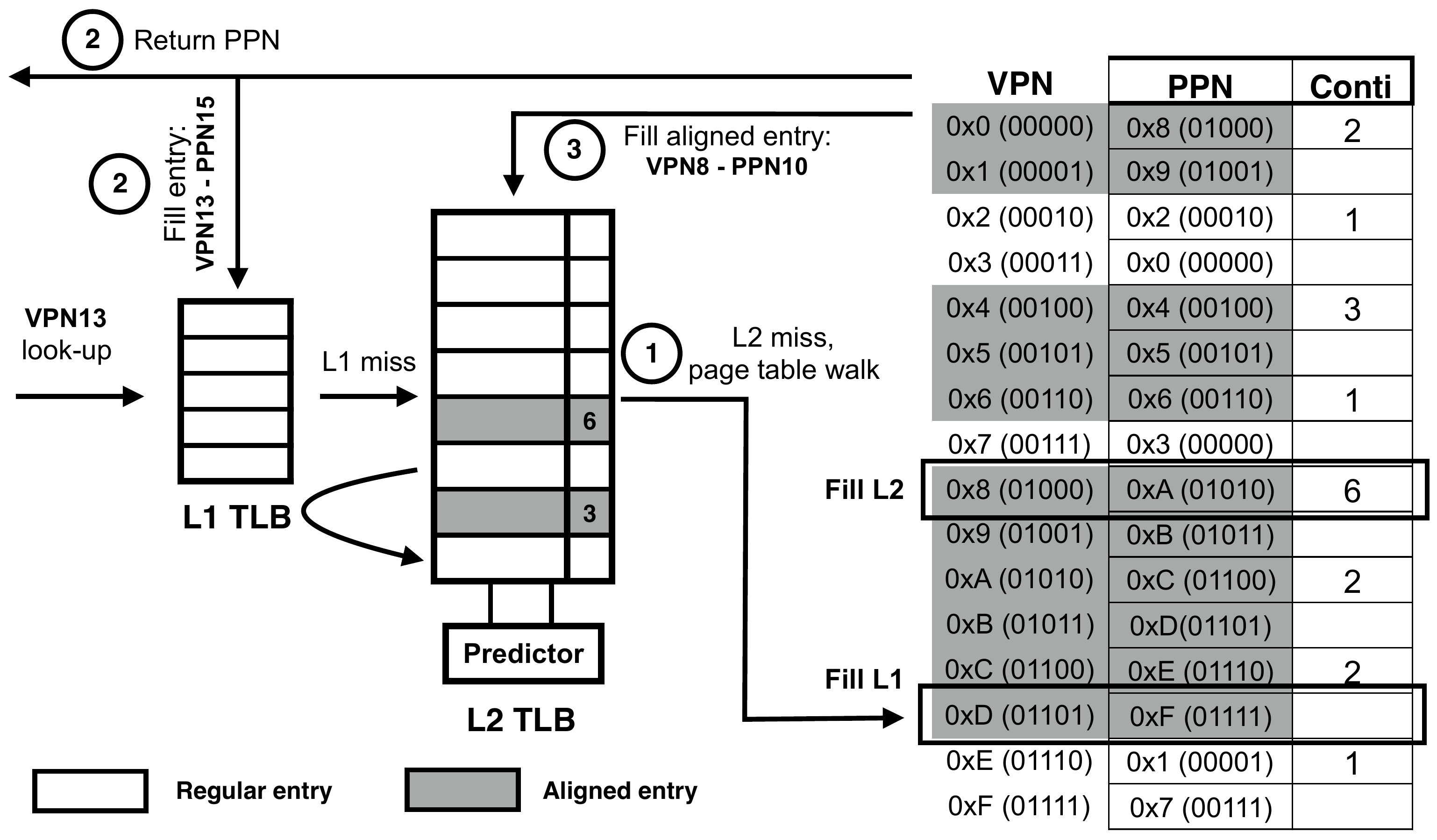}
 \caption{The \textit{fill} flow after page table walk.} 
 \label{fig:pt}
 %\vspace{-0.4cm}
\end{figure}

\para{TLB Fill.} On a L2 TLB miss, a page table walk occurs. Due to the urgency of execution, the PPN of required VPN is fetched and is delivered to L1 TLB and the core directly. \textit{In the background}, all the corresponding aligned entries identified by $\mathbf{K}$ need to be checked for looking up the one containing the maximal contiguity. The optimal aligned entry will be inserted into L2 TLB instead of regular entry because an aligned entry can complete translating all VPNs it covered in addition itself in the L2 look-up. 

Algorithm \ref{alg:fill} presents the workflow of \textit{L2 TLB fill}. Let VPN be the requested virtual address. For each $k \in \mathbf{K}$, the entry of VPN$_k$, the $k$-bit aligned VPN, is fetched and checked if VPN is covered by the entry's contiguity (lines 2-5). As we intend to find one with maximal contiguity, we search the aligned entry complying with the descending order of $\mathbf{K}$ (line 1). It is a guarantee that if $k$-bit aligned entry covers VPN, then the \%-bit aligned entries where \% $> k$ must cover VPN too. Therefore, the aligned entry is returned once the contiguity matches (lines 5-7). If all aligned entries fail to cover VPN, the matching entry of VPN is inserted to L2 TLB (line 8-10). In the worst cases,  OS requires  $|\mathbf{K}|$ times checks in order to find the optimal entry inserting L2 TLB. Note that these fetches and checks are no longer in the critical path of core execution.

\begin{algorithm}[ht]
\caption{ \textit{L2 TLB fill}} \label{alg:fill}
\begin{algorithmic}[1]
\Require  
$VPN$, $\mathbf{K}$   \ \# \textit{ $\mathbf{K}$ is set of natural numbers returned by Algorithm 3.}
\State sort $\mathbf{K}$ in descending order
\For{each $k \in \mathbf{K}$} \ \# \textit{start from the first element}
\State VPN$_k$ $\leftarrow k$-bit aligned(VPN) \  \# \textit{clear out $k$ LSB bits of VPN}
\State Entry $\leftarrow$ PageTable(VPN$_k$)   \  \# \textit{fetch the entry of  VPN$_k$.}
\If {Entry.contiguity $\geq$ (VPN $-$ VPN$_k$)}
\State insert Entry to L2 TLB
\State return
\EndIf
\EndFor
\State Entry $\leftarrow$ PageTable(VPN)
\State insert Entry to L2 TLB
\State return
\end{algorithmic}  
\end{algorithm}

 Figure \ref{fig:pt} shows an example of L2 TLB fill. TLB needs to translate VPN 13 to its PPN. For both L1 and L2 misses, a page table walk triggers. After sending PPN 15 to CPU and the mapping VPN 13 - PPN 15 to L1 TLB, OS decides which aligned entry to be filled to L2 TLB by Algorithm \ref{alg:fill}. VPN 10 and VPN 8 are the 2-bit and  3-bit aligned VPN respectively. Because VPN 8 has larger coverage, the 3-bit aligned entry is inserted into L2 TLB other than 2-bit. Analogically, 2-bit and even 1-bit will be considered if the involved contiguity chunk is smaller. Even though the sizes of contiguity chunks are various, OS can assign an optimal alignment for each chunk.

\begin{figure}
\centering
 \includegraphics[width=1.0\columnwidth]{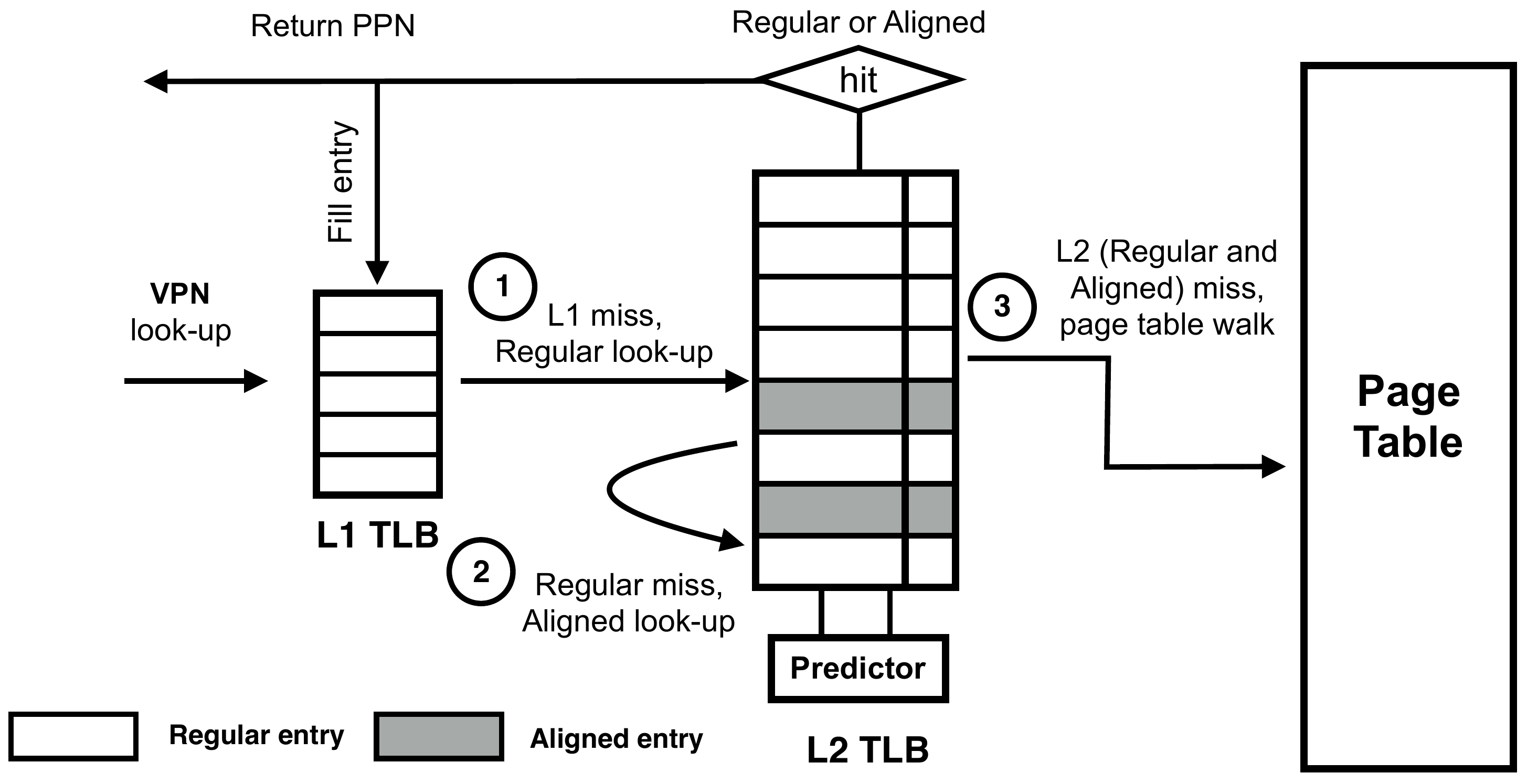}
  \vspace{-0.2cm}
 \caption{L2 TLB lookup: Regular lookup and Aligned lookup.} 
 \label{fig:lookup}
 %\vspace{-0.4cm}
\end{figure}

\para{ L2 TLB Look-up.}
Applying \textit{L2 TLB fill}  algorithm, L2 TLB includes both regular entries and aligned entries with stored contiguity. On a L1 TLB miss, L2 TLB first triggers the \textit{regular look-up}. The translation is done if it is a hit, using the physical paged number stored in the TLB entry. If a regular miss occurs, L2 TLB will starts \textit{aligned look-up}. 

$\mathbf{K}$ maintains all possible alignments that the requested VPNs work on, but we do not know which alignment can identify the corresponding aligned entry and complete the translation. One simple choice is  sequentially looking up all alignments of $\mathbf{K}$, as shown in Algorithm \ref{alg:look}. Given $k \in \mathbf{K}$, the $k$-bit aligned VPN, VPN$_k$,  is calculated by clearing out $k$ LSB bits of VPN (line 2). If an entry is found by locating VPN$_K$ and the value of the stored contiguity is greater than the difference between VPN and VPN$_k$ (contiguity matches), the PPN is finalized by the sum of its stored PPN and the difference (VPN - VPN$_k$) (lines 3-7).

\begin{figure}
\centering
 \includegraphics[width=1.0\columnwidth]{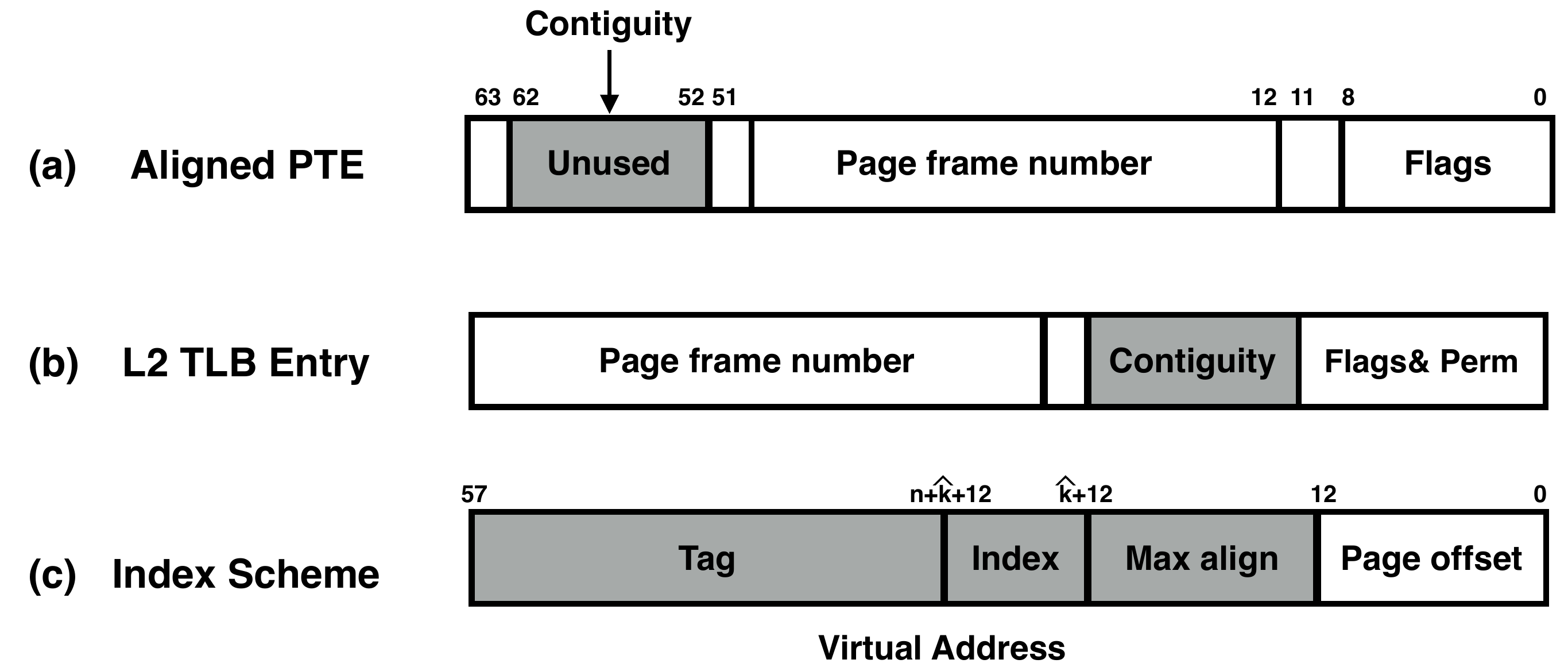}
  \vspace{-0.2cm}
 \caption{Modified entry and index scheme.} 
 \label{fig:entry}
 \vspace{-0.2cm}
\end{figure}

The indexing scheme of set-associative L2 TLB requires a few modifications, as shown in Figure \ref{fig:entry}. Let $\hat{k}$ be the maximum of $\mathbf{K}$. Virtual address $[12 : \hat{k}+12)$  denotes the bits $\hat{k}$-bit alignment with zero. To make full use of all TLB sets for aligned entries,  [$\hat{k}$+12 : $\hat{k}$+12+N) bits of the virtual address are used as the index bits, where N denotes the log2 (the number of sets).

\para{Speculation for Aligned Look-up.}
For an aligned look-up, the TLB needs to be looked up for $|\mathbf{K}|$ times in the worst cases. Therefore, as $\mathbf{K}$ expands, the overheads of aligned look-up may increase linearly. 
To reduce this cost, We expect the aligned look-up can be finished in the first look-up rather than multiple times. Thus, we propose to predict the exact alignment in $\mathbf{K}$, which directly makes the requested VPN succeed in translation.

Memory requests exhibit some predictable behavior, which has been widely applied on not only cache prefetching but also TLB prefetching \cite{prefetch1,prefetch2} such as sequential Prefetching and Stride Prefetching. Heuristically, we consider these predictable behavior in the aligned entries. An aligned entry is coalesced by a range of contiguous PTEs.

%%%%too complex

If the requested VPN is located at one page of the range, it is high likely that the following address translations also belong to the pages of this range due to the spatial locality and temporal locality.
Therefore, we speculate that these consecutive requests share one aligned entry (i.e., \textit{a same alignment of $\mathbf{K}$}). 

Based on this observation, we add a 4-bit predictor to the L2 TLB (assume the upper bound of $|\mathbf{K}$| is 8), recording the latest used alignments. In the aligned look-up, the alignment kept by the predictor will be used to look up the aligned entry of requested VPN at the first place. The other remaining alignments are used sequentially if the first look-up fails (prediction fails). In our evaluation, the accuracies of  the predictor are 92.8\%, 91.6\%, and 91.2\% for all aligned hits, when $|\mathbf{K}|$ are 2, 3, and 4 respectively.

\begin{algorithm}[ht]
\caption{ \textit{L2 TLB Aligned Look-up}} \label{alg:look}
\begin{algorithmic}[1]
\Require VPN, $\mathbf{K}$ 
\Ensure PPN  \  \# \textit{ mapped by VPN.}
\For{each $k \in \mathbf{K}$}
\State VPN$_k$ $\leftarrow k$-bit aligned(VPN) 
\State Entry $\leftarrow$ PageTable(VPN$_k$)
\If{Entry is found}    \  
\If {Entry.contiguity $\geq$ (VPN $-$ VPN$_k$)}
\State PPN $\leftarrow$ Entry.PPN + ( VPN $-$ VPN$_k$)
\State return PPN     \  \# \textit{Translation is completed.}
\EndIf
\EndIf
\EndFor
\State return \  \# \textit{Translation is failed.}
\end{algorithmic}  
\end{algorithm}

To sum up,  L2 TLB with $\mathbf{K}$-bit aligned entries boosts the translation coverage of TLB by mining contiguous pages of page table,  decreasing the miss rate of  TLB (regular and aligned look-up). 
%%% two calculation ??
%it requires two modifications: one is choosing the optimal entry to be insert in the \textit{fill}; another one is the aligned look-up.
 %Nonetheless, 
Meanwhile,  we try to hide the translation latency it produced and make minor changes to MMU. First, in the \textit{fill} process, as the translated PPN and required PTE are first delivered to the core and L1 TLB, the selection (Algorithm \ref{alg:fill}) is no longer on the critical path of core execution. Second, we add a predictor to L2 TLB, boosting the potential of finishing aligned hit in one look-up, and thus the overhead of aligned look-up is significantly decreased.

\subsection{Determining $\mathbf{K}$}

\begin{table}[htbp]
\caption{Size Range Table}
\begin{center}
\begin{tabular}{|c|c|}
\hline
Size Range & \%-bit  Alignment \\
\hline
$ 2 \sim  16$ & 4  \\
$17 \sim 64$ & 6 \\
$65 \sim 128$ & 7 \\
$129 \sim 256$ & 8 \\
$257 \sim 512$ & 9 \\
$513 \sim 1024$ & 10 \\
 $ >  1024$ & 11 \\
\hline
\end{tabular}
\label{tab1}
\end{center}
\end{table}
In this section, we propose an algorithm to determine $\mathbf{K}$ based on the distribution of contiguity chunks in page table. In the process initial phase, many processes allocate the majority of memory they use for the rest of execution \cite{efficient,anchortlb}. Once the initial memory allocation phase is stabilized, the status of memory mapping is used to ensure $\mathbf{K}$.

As the illustration in Section 3, in many real-world applications, memory mappings are awash with diverse sizes of contiguity chunks.
The distribution of aligned entries determined by the optimal $\mathbf{K}$ should most fit the distribution of contiguity chunks, i.e., every contiguity of chunks covered by its matching aligned entry. 
%The certain size of contiguity chunk has the most suitable aligned entry. %% can't understand
For example, a $6$-bit aligned PTE enables a contiguity chunk of size 64 to be completely covered by its contiguity. The memory contiguity information is assessed by the contiguity histogram \cite{anchortlb} maintained by OS. Contiguity histogram records the number of contiguity chunks of varying sizes allocated to the process.

First, we define the \textit{size ranges} of contiguity chunk and their matching alignment by a heuristic approximation, as the presentation in Table \ref{tab1}. To achieve the optimum, $\mathbf{K}$ should mirror the most frequently occurring sizes of contiguity chunks which are counted by contiguity histogram. Accordingly, we greedily choose the alignment obeying the frequency of size range of contiguity chunk.

Algorithm \ref{alg:det} shows the high level description for determining $\mathbf{K}$. 
In lines 1-9, we iterate through all contiguity chunks allocated to this process, and assign a matching aligned entry to each chunk in line with Table \ref{tab1} (line 5). Assuming each chunk is covered by its corresponding aligned entry, we calculate the coverage for each type (denoted by $k$) of alignment respectively, by accumulating the contiguous pages in the matching chunks (lines 6-9). The coverage can be thought of as the weighting of each type of alignment. Next, we choose the alignment obeying the descending order of the coverage (line 11-13). 
Actually, computing cost increases as $|\mathbf{K}|$ increases, we expect to use the least types of alignment to cover the most sizes of contiguity chunks.
Thus, we provide $\theta$ (lines 15-16). 
In the evaluation, we set $\theta$ as $0.9$, which indicates that $\mathbf{K}$ does not add when the aligned entries ensured by $\mathbf{K}$ covers more than $90$ \% of contiguous pages. 
$\psi$ is to give a upper bound of $|\mathbf{K}|$, which is set as $4$ in our experiment. For example, if the memory mapping is filled with the contiguity chunks of size 16 and 128 that cover more than 90\% of contiguous pages, $\mathbf{K} = \{4, 7\}$ will be returned by Algorithm \ref{alg:det}.   

\para{Stability of $\mathbf{K}$}.
Memory mapping may change significantly during the execution as the allocation and deallocation of memory, bringing challenges to existing TLB coalescing techniques. In the proposed scheme, this impact is mitigated conspicuously. As the introduction of $\mathbf{K}$-bit aligned PTEs, the chances of coalescing the varying sizes of contiguity chunk size into an aligned entry increase, and hence the tolerance for the change of distribution of contiguity chunk is reinforced. To check the influence of the memory mapping change to $\mathbf{K}$,  we record the varying of contiguity chunks during the execution in real machine, and find that the number and size of contiguity chunk rarely change after the memory allocation is completed, which is similar to the fact observed by \cite{anchortlb}
Therefore, $\mathbf{K}$ is unneeded to change in many cases. In the evaluation, we execute Algorithm \ref{alg:det} to update $\mathbf{K}$ every five billion instructions.

\begin{algorithm}[ht]
\caption{ \textit{Determining $\mathbf{K}$}} \label{alg:det}
\begin{algorithmic}[1]
\Require contiguity\_histogram, $f()$ \ \#  \textit{contiguity\_histogram is a list of (size, freq) pairs. E.g., (16, 33) denotes that the contiguity chunk of size 16 occurs 33 times in the memory mapping. $f()$ is the mapping function of Table \ref{tab1}: input Size, output Alignment.}  
\Require alignment\_weight  \ \# \textit {alignment\_weight is a list of (Key, Value) pairs.}
\Ensure $\mathbf{K}$  
\State $\mathbf{K} \leftarrow \emptyset$, coverage $\leftarrow 0$, total\_contiguity $\leftarrow 0$
\For {each (size, freq) in contiguity\_histogram}
\State coverage $\leftarrow$ size $\times $ freq
\State total\_contiguity $\leftarrow$ total\_contiguity + coverage
\State $k \leftarrow f(size)$
\If {$k$ \textbf{not in} alignment\_weight}
\State alignment\_weight $\leftarrow$ ($k$, coverage)
\Else
\State alignment\_weight[$k$] $\leftarrow $ alignment\_weight[$k$] + coverage 
\EndIf
\EndFor
\State sum\_coverage $\leftarrow  0$  
\State sort alignment\_weight in the descending order according to Value
\For {each ($k$, coverage) in alignment\_num} \ \# \textit{start from the first pair}
\State $\mathbf{K} \leftarrow \mathbf{K} \ \cup $ \{Key\}
\State sum\_coverage $\leftarrow$ sum\_coverage + coverage
\If {sum\_coverage $>$ total\_contiguity $\times \theta$  \ \# \textit{ $\theta \in (0, 1]$, $\theta$ denote the percentage of total contiguity that aligned entries are required to cover }} 
\State break
\EndIf
\If {$|\mathbf{K}| > \psi $ \ \# \textit{ $\psi$ is the maximal size of $\mathbf{K}$ }}
\State break
\EndIf
\EndFor
\State return $\mathbf{K}$

\end{algorithmic}  
\end{algorithm}

\subsection{ OS Implication}

As the existing hybrid coalescing approach\cite{anchortlb}, the contiguity information of aligned entries is maintained by OS. There are two minor changes required on OS: updating memory mapping and initialization of $\mathbf{K}$-bit aligned entries.

\para{Updating Memory Mapping.} During the execution of process, a physical memory frame can be allocated, relocated, or deallocated by OS. Once a memory frame is updated, the contiguity values of the related aligned entries confirmed by $\mathbf{K}$ also need to be updated. After updating the page table entries and aligned entries, OS triggers a conventional TLB shootdown, invalidating all entries from the TLBs of all cores.

\para{Initialization of $\mathbf{K}$-bit Aligned Entries.}  With the stabilization of initial memory allocation, $\mathbf{K}$ is determined by Algorithm \ref{alg:det}, and accordingly the aligned entries that record how many following pages are contiguous are confirmed. To calculate the contiguity stored in all aligned entries, OS need to traverse the entire memory mapping once. First, considering updating one type of alignment, let $\mathbf{K} = \{ \underline{k} \}$ and $N$ be the total number of pages in memory mapping. Next, $N/2^{\underline{k}}$ aligned entries need to be updated.
Now, append more types of alignment to $\mathbf{K}$, $\mathbf{K} = \{\underline{k}, a, b, c ...\}$, and assume $\underline{k}$ is the minimum of $\mathbf{K}$.
Based on the proposed design of page table, the number of aligned entries need to be updated also is $N/2^{\underline{k}}$. Therefore, the computing cost of initialization of $\mathbf{K}$-bit aligned entries is nearly identical to updating $\underline{k}$-bit aligned entries where $\underline{k}$ is the minimum of $\mathbf{K}$.

We conduct experiments to collect the overhead of initialing aligned entries with different $\mathbf{K}$.  When $\mathbf{K} = \{4\}, \{4,5\},$ and $\{4,5,6,7,8,9\}$,  the consuming time of traversing the entire page table are 162ms, 162.4ms, 163ms respectively, when a process occupies 18 GB memory. Furthermore, the consuming time is 354.3ms, 78.8ms, and 3.2ms respectively if $\mathbf{K} = \{3,4\}, \{5,6\}, $and$ \{8,9\}$. In practice, $\mathbf{K}$ can be set default values as a process is created. Consequently, the contiguity values of aligned entries are calculated when the memory is allocating, resulting in that the cost of initialization of $\mathbf{K}$ determined by Algorithm \ref{alg:det} slumps because many contiguity values of aligned entries do not change.
After updating $\mathbf{K}$-bit aligned entries, the TLBs need to be invalidated, which is a relatively minor cost based on the fact the native Linux kernel for x86 flushes the TLB on context switches \cite{anchortlb}.
 
\para{Permission and Page Sharing.}
Pages may have different r/w/x permissions, which impede coalescing contiguous pages. Although the proposed scheme can simply treat a page with a different permission as the non-contiguous page, \cite{efficient} has found that permissions are commonly homogeneous in contiguity chunks and thus the impact of permission is minor. For page sharing across processes, aligned entries record the contiguity for the corresponding page table of each process, achieving the TLB coalescing.

\subsection{Future Work}
Aligned lookup finds the corresponding aligned entry of request VPN if it exists and completes the address translation by a simple calculation, preventing the trigger of page table walk. Admittedly, aligned lookup is on the execution path of CPU and produces additional latencies for an L2 TLB miss (both regular and aligned miss), because the page table walk starts after the aligned lookup. Although aligned lookup significantly reduces the TLB misses, this caused latency for an L2 TLB miss still can be hidden. Using speculation in Sec 5.2 let near 90 \% aligned hit be finished in one lookup, and therefore we propose to \textit{start the second lookup in parallel with page table walk} if the first lookup misses. Therefore, the second or subsequent lookups do not add additional latency for an L2 TLB miss. This design requires the prefetch buffer and logic to deal with the conflict that aligned hit happens after starting the page table walk.  We believe that such approach would further improve the MMU performance, and we leave this as future work.

\section{Experiment}
 
This section describes the configuration used in the experiments and the evaluation for the proposed approach, denoted by $\mathbf{K}$ Aligned, and four baselines in comparison. We simulated the works on a trace-based simulator and generated the memory access trace of 10 billion instructions using Pin binary instrumentation tool\cite{luk2005pin}.

\subsection{Methodology}

\para{Workloads.}We used benchmarks from SPEC CPU 2006\cite{spec2006} and graph500 and gups\cite{efficient}. The working set sizes of graph500 and gups are set to 8GB.

\para{Used Mappings.} We captured virtual-physical address mappings using the pagemap\cite{pagemap} interface provided by Linux. The first mapping used in experiments is generated on a Linux v4.16 machine using the default \textit{demand paging} and set THP on. In addition to this real mapping, we synthesized four mappings to observe the effects of contiguity types for previous methods: small contiguity, medium contiguity, large contiguity, and mixed contiguity. Table \ref{tab:syn} presents the distribution of contiguity chunks for each contiguity type, where the sizes of chunks are randomly formed from the given range. For mixed contiguity, we select the contiguity chunks size ranges obeying the weight of each size range to reveal the disadvantages of existing approaches and check the effectiveness of proposed scheme.

\para{Comparisons.} We compared $\mathbf{K}$ Aligned   with the five following prior coverage improvement techniques: THP\cite{THP},  COLT\cite{colt}, Cluster\cite{cluster}, RMM\cite{RMM}, and Anchor\cite{anchortlb}. The baseline configuration is the default TLB of Linux without any modification and THP is an implementation of huge page (2MB) in linux.   
Both Cluster and RMM require extra hardware to support coalescing, where the configurations is as used by \cite{anchortlb}.   To obtain optimal performance, Anchor has two implementation schemes: \textit{dynamic} and \textit{static}. The former one is using the dynamic distance selection algorithm to determine anchor distance, contrary to the latter one that exhaustively tries all possible anchor distance and ends up with the optimal performance. 

To evaluate the performance effects of $\mathbf{K}$ of proposed method, we use three modes of proposed approach in comparisons by varying $\mathbf{K}$: $|\mathbf{K}| = 2$ Aligned, $|\mathbf{K}| = 3$ Aligned, and  $|\mathbf{K}| = 4$ Aligned.

\para{TLB Parameters.} Table \ref{tab:con} exhibits the TLB configurations for all methods. The L1 TLB configuration is the same for all approaches. The L2 TLB capacity is set to 1024 entries. In addition to regular L2 TLB, Cluster possesses an additional clustered TLB  and RMM holds a 32-entry fully associative range TLB. Anchor and $\mathbf{K}$ Aligned have the same configuration. 
Note that all regular TLBs support both 4KB and 2MB page sizes.

\para{Evaluation Metrics.} We used TLB misses and CPI to evaluate the performance of each method. The lower part of Table \ref{tab:con} gives the latency parameters used in the experiments. L1 TLB is accessed in parallel with the cache access and thus the L1 TLB access latency is hidden \cite{efficient}. Note that the latency of aligned lookup is calculated by accumulating the required L2 TLB lookups.

\vspace{-0.5em}
\begin{table}[h]
\footnotesize
\caption{ TLB configuration used for evaluation}
\vspace{-0.7em}
\begin{tabular}{c|l}
\toprule
 Schemes &  TLB Configuration\\
\toprule
\multirow{2}{*}{Common L1 TLB} & 4KB: 64 entries, 4-way\\
&   2MB: 32 entries, 4-way \\
\midrule
Base \& THP & 1024 entries, 8-way\\
\midrule 
COLT &  1024 entries, 8-way \\
\midrule 
\multirow{2}{*}{Cluster}& Regular TLB: 768 entries \\
& 6 way Cluster-8: 320 entries, 5-way  \\
\midrule 
\multirow{2}{*}{RMM} & Baseline L2 TLB\\
  & 32 entries, fully associative   \\
\midrule 
Anchor &  1024 entries, 8-way   \\
\midrule
\textbf{K} Aligned & 1024 entries, 8-way \\
\bottomrule
\multirow{3}{*}{Latency} & 7 cycles, L2 hit \cite{l2cycle}\\
& 8 cycles, cluster/RMM/anchor/aligned hit \cite{anchortlb}\\
& 50 cycles, page table walk \cite{efficient}  \\
\bottomrule
\end{tabular}
\centering
\label{tab:con}
\vspace{-1em}
\end{table}

\vspace{-0.5em}
\begin{table}[h]
\footnotesize
\caption{ Configurations of synthetic mappings}\label{tab:syn}
\vspace{-0.7em}
\begin{tabular}{c|l}
\toprule
& Contiguity Chunk (4KB page)\\
\toprule
Small Contiguity  &  1-63 pages  \\
\midrule
Medium Contiguity & 64-511 pages \\
\midrule
Large Contiguity  & 512-1024 pages  \\
\midrule
mixd contiguity & 0.4 Small + 0.4 Medium + 0.2 Large \\
\bottomrule
\end{tabular}
\centering
\vspace{-1em}
\end{table}

\begin{figure*}[ht]
\centering
 \includegraphics[width=1.0\textwidth]{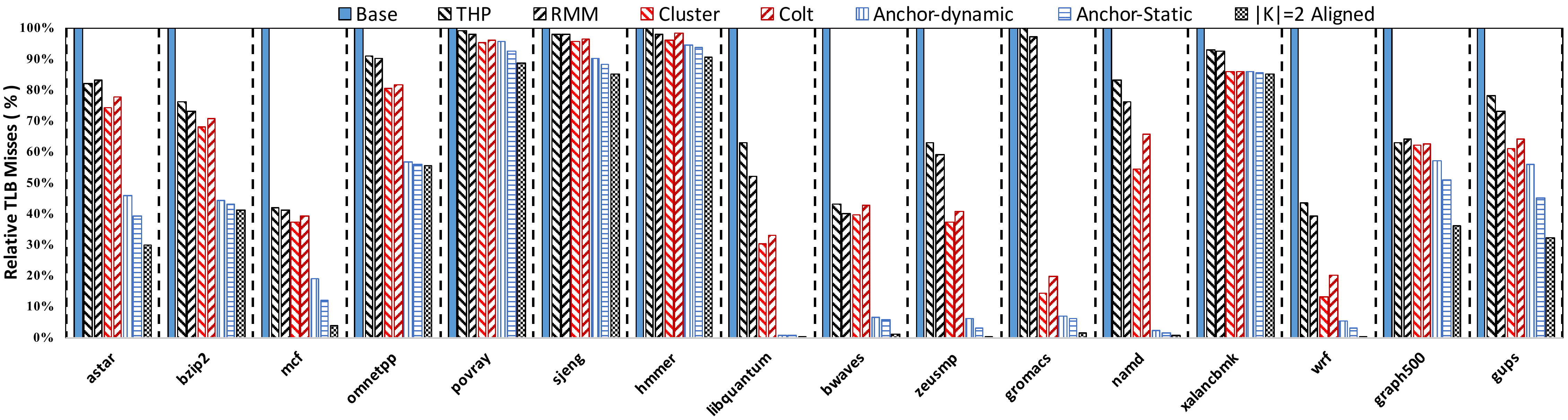}
  \vspace{-0.4cm}
 \caption{Relative misses of all compared approaches for \textit{demand} mapping.} 
 \label{fig:pt1}
 \vspace{-0.2cm}
\end{figure*}

\begin{figure*}[ht]
\centering
 \includegraphics[width=1.0\textwidth]{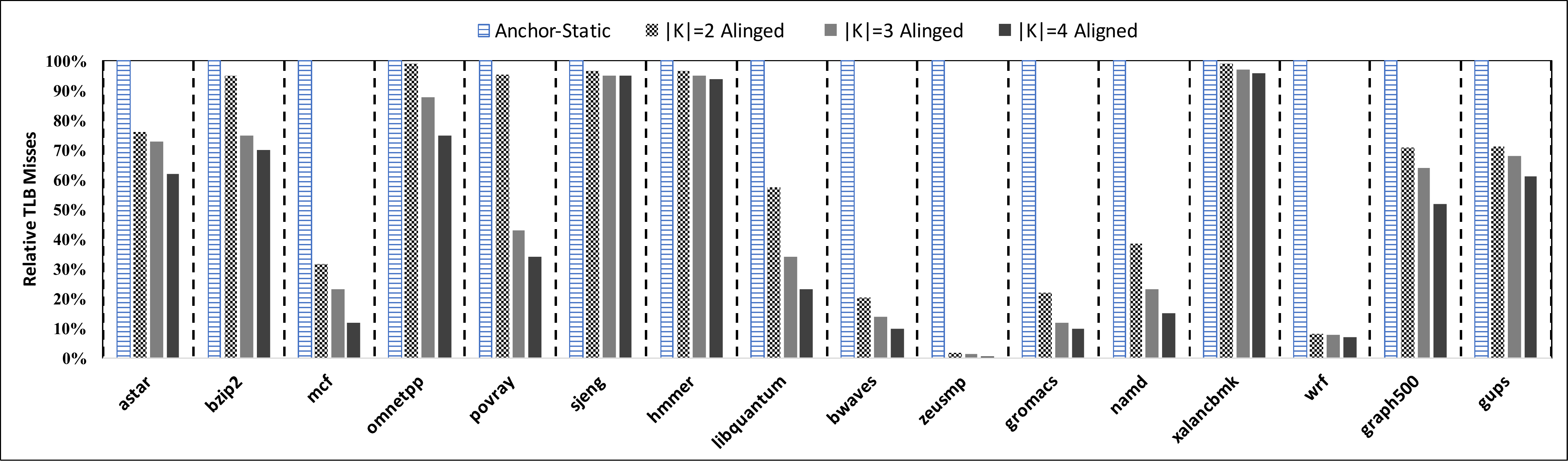}
  \vspace{-0.4cm}
 \caption{Relative misses of varying $|\mathbf{K}|$ for \textit{demand} mapping.} 
 \label{fig:pt2}
 \vspace{-0.2cm}
\end{figure*}

\subsection{Results}

In this section, we presents the details of evaluation by exhibiting the relative TLB misses with the real mapping (demand mapping) for all completing approaches in all benchmarks. Because of the constrained space, we only show the mean TLB reductions while using synthetic mappings.

\para{Real Mapping.}
Figure \ref{fig:pt1} plots the relatives TLB misses normalized to the \textit{Base}.
For $\mathbf{K}=2$ Aligned, we set $ \psi = 2$ and the maximal number of types of alignments it supports is $2$. 

In summary, our method outperforms all other prior translation coverage techniques by reducing the 69.2\% TLB misses on average relative to Base. It maximizes the TLB coverage by filling the coalesced entries that are deliberately selected for the varying sizes of contiguity chunks. In contrast, other approaches only provides one fixed sizes of container for all types of contiguity chunks. Thus, while encountering the complicated contiguity distribution, they become not effective. 

The TLB misses reduced by $|\mathbf{K}|=2$ Aligned varies on each benchmark, which is related to the contiguity of corresponding benchmark, as shown in Figure \ref{fig:cc}. The more contiguity chunks allocated by OS, the larger TLB coverage can be improved by coalescing techniques, and the more various all approaches's performances are. For example, more than 9K contiguity chunks are allocated for the application \emph{mcf} in our experiments, and these chunks are disparate in terms of size. In mcf, THP alone can effectively reduce the TLB misses by 58\%, using 2MB pages to coalesce large contiguity chunks. With the support of THP,  COLT and Cluster benefit from small contiguity via HW coalescing logic, in addition to large contiguity, achieving 61\% and 63\% misses reduction respectively. As RMM also focuses on large contiguity, it reaches the similar TLB miss rate with THP, decreasing TLB misses by 59\%. Anchor-Static decrease the TLB misses by 88\%, which can be thought of the best performance \cite{anchortlb} can achieve. However, apart from large contiguity (captured by THP), it can dynamically change suit only one type of contiguity (small contiguity in mcf).  $|\mathbf{K}|=2$ Aligned is beyond Anchor-Static by supporting two types of coalesced entries and finally suit two types of contiguity \textit{simultaneously} (both small and medium contiguity). Unsurprisingly, $|\mathbf{K}|$=2 Aligned achieved the best performance, reducing TLB misses over 96\%.

Table \ref{tab:avgxx} exhibits the average TLB misses of $\mathbf{K}$ Aligned and prior approaches, relative to Base.  The best prior method, Anchor-Static, achieves  42\% relative misses. The proposed method where $|\mathbf{K}|=2$ can further decrease the TLB miss by 27\% over Anchor, using two types of alignments.

\begin{table*}[th]
\footnotesize
\caption{ The relative misses on average of all approaches for each memory mapping.} \label{tab:avgxx}
\vspace{-0.7em}
\begin{tabular}{c|c|cccccc|ccc}
\toprule
 & & Base &THP \cite{THP}  &RMM \cite{RMM} &COLT \cite{colt}& Cluster\cite{cluster} &Anchor-Static\cite{anchortlb} & $|\mathbf{K}$|=2 Aligned & $|\mathbf{K}$|=3 & $|\mathbf{K}$|=4 \\
\midrule
Real Mapping & Demand   & 100\% & 80\% & 78\% & 67.6\% &70.7\% &42\% &30.8\% & 21.7\% & 18.9\% \\   
\midrule
\multirow{4}{*}{Synth Mapping} & Small & 100\% & 100\% & 99.2\% & 60.5\% &55\% &45.3\% &35.9\%  & 33.4\% & 31.2\%  \\
&Medium  & 100\% & 100\% & 99.3\% & 56.1\% & 52.3\% & 33.4\% &25\% & 20.4\% &  17.4\%\\
&High  &100\% & 45.6\% & 45.1\% & 34\% & 38.2\% &10.3\% &6.4\% & 4.3\% & 3.9 \% \\
&Mixed  &100\% & 81.2\% & 72.4\% & 56.3\% & 53.2\%  & 60.5\%  &25\% & 13.2\% & 5.6\% \\
\bottomrule
\end{tabular}
\centering
\end{table*}

\para{Synthetic Mappings.}
We used four synthetic mappings to simulate the possible contiguity types in real applications. 
Table \ref{tab:avgxx} shows the average results of each method on each mapping.

The first three mappings is to describe the abstract contiguities. In detail, the sizes of allocated contiguity chunk are restricted to a narrow range, but these abstract contiguities are the ideal contiguity distribution for prior techniques. In the small contiguity, COLT and Cluster succeed in coalescing these small contiguity chunks, decreasing 60.5\% and 55\% relative TLB misses. However, as the contiguity chunks are less than 2MB, THP and RMM showed the feeble performance. 
On the other hand, the large contiguity provides large contiguity chunks for THP and RMM, almost eliminating the TLB misses. Nevertheless, the COLT and COLT didn't further reduce the TLB misses as the augmentation of contiguity.
Anchor can change its anchor distance to suit a type of contiguity and therefore it maintains effectiveness on first three abstract mappings. $|\mathbf{K}|=2$ Aligned is capable to suit two types of contiguity dynamically, and it performs as well as we expected on the first three abstract mappings.

The mixed mapping is derived from the observation of contiguity in real applications. the prior techniques are no longer very effectiveness with these diverse contiguity chunks. Our hybrid coalescing methods are more effective to detect the available contiguity, excelling the other schemes. Compared with Anchor, $|\mathbf{K}|=2$ Aligned further reduces  58\% TLB misses.

\para{Coverage Improvement.}      
At every billion instruction boundary, we accessed the L2 TLB to record the TLB translation coverage. In the baseline configuration, the coverage is the number of inserted entries of TLB. For COLT, Anchor and $\mathbf{K}$ Aligned, the TLB coverage is the number of inserted entries plus the sum of contiguity values of every coalesced entries. For example, the coverage of an aligned entry in Figure \ref{fig:pt} is itself (1) plus the number of following contiguous pages it covered (contiguity). As the RMM and Cluster need additional TLB HW, we did not count them in this statistics.

Table \ref{tab:avg} gives the average coverage of each scheme of three times record. The TLB of COLT is filled with regular entries and coalesced entries. Due to the restriction of HW coalescing, the modified entry is able to store up to 8 following contiguous pages. Therefore the coverage of COLT is larger than Base (1024, the number of regular entries). In contrast, Anchor and $\mathbf{K}$ Aligned compress contiguity chunks into modified entries in the page table by OS, almost erasing the upper bound of the number of entries for coalescing. Unsurprisingly, Anchor has larger coverage across all applications. Unfortunately, with diverse sizes of contiguity chunks, the efficiency for coalescing of Anchor will be decreased.  $\mathbf{K}=2$ Aligned provides more types of coalesced entries to mitigate this issue, and the coverage is increasing as the $|\mathbf{K}|$ grows, which explains its unsurpassed performance for TLB misses reduction.

\begin{table}[th]
\footnotesize
\caption{ The comparison of relativeTLB translation coverage (covered PTEs)} \label{tab:avg}
\vspace{-0.7em}
\begin{tabular}{c|ccc|c}
\toprule
 & Base (1024) & COLT & Anchor-Static & $|\mathbf{K}|=2$ Aligned\\
\midrule
astar & 1 & 5.34 & 6.56 & 7.23\\
bzip2 &1 & 4.34 & 8.24 & 9.6\\
mcf & 1 & 6.2 & 23.44 & 34.2\\
omnetpp & 1 & 1.34& 3.21 & 4.58\\
povray & 1 & 1.78& 2.2 & 2.51\\
sjeng & 1 & 1.4& 1.89 & 1.98\\
hmmer & 1 & 1.39& 1.58 & 1.68\\
libquantum & 1 & 4.31& 16.47 & 23.35\\
bwaves & 1 &2.82& 7.34 & 15.4\\
zeusmp & 1 & 1.98 & 7.85 & 13.48\\
gromacs & 1 & 4.32& 9.45 & 15.79\\
namd & 1 & 2.34& 9.84 & 13.24\\
xalancbmk & 1 & 1.34& 1.42 & 1.56\\
wrf & 1 & 3.23 & 8.57 & 10.78\\
graph500 & 1 & 2.45 & 4.56 & 5.98\\
gups & 1 & 3.41& 5.43 & 8.23\\
\bottomrule
\end{tabular}
\centering
\end{table}

\para{Effectiveness of K.}
Determining $\mathbf{K}$ is the key of our scheme. We implemented  $\mathbf{K}$ Aligned using Algorithm 3 and made evaluations when $\psi = 2, 3,$ and $4$ respectively. As more types coalesced entries are provided, the coverage of TLB must be larger in face of the same mapping with the increasing of $|\mathbf{K}|$, on account of Algorithm 1 that chooses the optimal aligned entries for coalescing. 
Figure \ref{fig:pt2} shows TLB misses of varying $|\mathbf{K}|$ Aligned, relative to Anchor. In summary, the more types of contiguity contained by the memory mapping, the more effective the bigger $|\mathbf{K}|$ should be.  Compared to $|\mathbf{K}|=2$ Aligned, $|\mathbf{K}|=3$ and $|\mathbf{K}|=4$ further reduce 30\% and 39\% misses respectively.

\para{Effectiveness of Predictor.}
Granted, the Aligned look-up of L2 TLB is on the execution path of CPU and thus the performance of TLB is sensitive to  the overheads of it. To accelerate the aligned lookup, we add a predictor beside L2 TLB to predict the exact alignment used the a Aligned hit. If the prediction is correct, the align look-up only needs one lookup of TLB, which is the same cost as COLT/Cluster/RMM/Anchor hit for an aligned hit.

Table \ref{tab:hit} shows the predictor's accuracy on all benchmarks with increasing $|\mathbf{K}|$. As the spatial locality of instruction accesses, the alignment used for translation in Aligned lookup exhibits predictable behaviors. We discover that a cluster of translation requests share the same alignment. Thus, we use a HW predictor to store the most recent used alignment and the aligned lookup start up with this alignment. Across all applications, the predictor achieved 94.3\% accuracy on average for $|\mathbf{K}|=2$ . When $|\mathbf{K}|$ increases, the predictor can maintain the accuracy because the spatial locality of a memory mapping does not change and the coverage of an aligned entry may be improved. Even $|\mathbf{K}| = 4$,  93.1\% aligned hit can be finished in one lookup. Thus, the proposed method can maintain effectiveness despite the growing $|\mathbf{K}|$.

\begin{table}[th]
\footnotesize
\caption{ The accuracy of predictor for Aligned Lookup. For an aligned hit, a prediction is correct if the address translation is completed in the first TLB lookup.} \vspace{-0.7em}
\begin{tabular}{c|ccc}
\toprule
   &$|\mathbf{K}|=2$ Aligned  & $|\mathbf{K}|=2$ Aligned & $|\mathbf{K}|=2$ Aligned  \\
\midrule
astar & 95.4\% & 94.3\% & 93.2\%  \\
bzip2 & 92.3\% & 93.4\% & 92.8\%  \\
mcf & 97.9\% & 97.5\% & 97.2\%  \\
omnetpp & 83.2\% & 78.3\% & 79.2\%  \\
povray & 89.4\% & 94.2\% & 94.1\%  \\
sjeng & 85.3\% & 84.8\% & 84.8\%  \\
hmmer & 90.3\% & 89.2\% & 89.1\%  \\
libquantum & 95.3\% & 95.6\% & 95.5\%  \\
bwaves & 98.2\% & 97.2\% & 97.3\%  \\
zeusmp & 98\% & 98.1\% & 98.1\%  \\
gromacs & 94.2\% & 93.4\% & 92.9\%  \\
namd & 93.4\% & 93\% & 93.2\%  \\
xalancbmk & 86.4\% & 86.4\% & 86.4\%  \\
wrf & 97.2\% & 97.6\% & 97.5\%  \\
graph500 & 92.3\% & 91.8\% & 91.7\%  \\
gups & 93.2\% & 93.0\% & 92.9\%  \\
\midrule
average &  94.3\% & 93.7\%  & 93.1\% \\
\bottomrule
\end{tabular}
\label{tab:hit}
\centering
\end{table}

\begin{figure}[ht]
\centering
 \includegraphics[width=1.0\columnwidth]{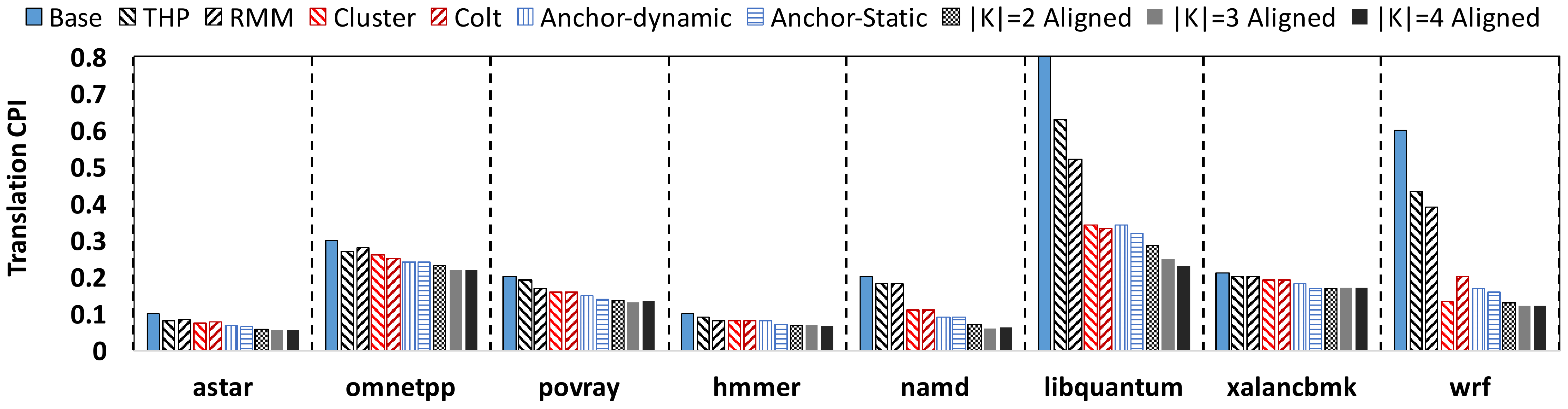}
  \vspace{-0.3cm}
 \caption{ CPI breakdown of translation overhead for \textit{demand} mapping} 
 \label{fig:pt31}
 \vspace{-0.2cm}
\end{figure}

\begin{figure}[ht]
\centering
 \includegraphics[width=1.0\columnwidth]{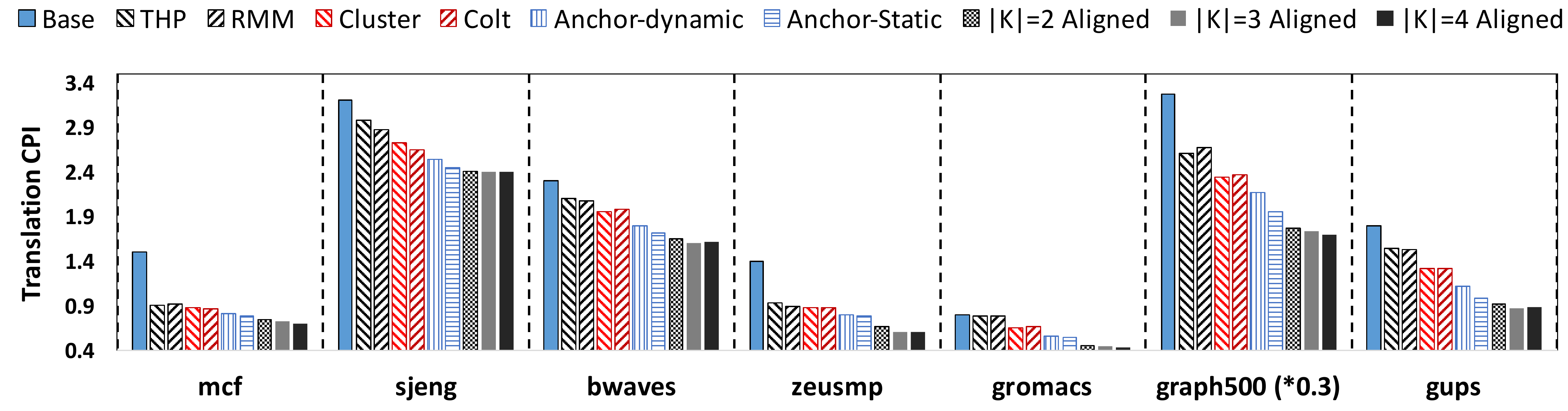}
  \vspace{-0.3cm}
 \caption{CPI breakdown of translation overhead for \textit{demand} mapping} 
 \label{fig:pt32}
 \vspace{-0.2cm}
\end{figure}

\para{Translation CPI.}
Figure \ref{fig:pt31} and \ref{fig:pt32}  show the breakdown of the cycles spent per instruction in the address translation. We estimate CPIs based on the latency units in the Table \ref{tab:con}. The overheads of aligned lookup is up to the number of lookup needed. If aligned lookup is completed in a lookup, then the latency produced is 8 cycles. Otherwise, add 7 cycles for each additional lookup. First, the results of CPIs is consistent with the TLB misses analysis for Figure \ref{fig:pt1} and \ref{fig:pt2}.
The proposed scheme outperforms the best prior approaches across all applications. Particularly, the performance improvements of $\mathbf{K}$ Aligned are significant on the applications with high contiguity. With $|\mathbf{K}|=2$, the CPI are 0.91, 0.74, and 5.92 for gups, mcf and graph500, respectively, for the real paging mappings. Compared to $|\mathbf{K}|=2$ Aligned, $|\mathbf{K}|=3$ and $|\mathbf{K}|=4$ are better or extremely similar, because of their further TLB miss reduction and the maintained efficiency of aligned lookup by the predictor.

\section{Related Work}

\para{Improving Coverage.} To reduce the overheads of virtual address translation, improving the translation coverage has become a primary approach\cite{huge2,efficient,THP2,NVM2,RMM,predict1,Efficientsynonym,cluster,colt,zhang2010enigma,anchortlb,DVM}. \cite{huge2} first proposed to use a single extended TLB entry to coalesce a subblock of pages in memory mapping, which paves a way for the HW coalescing techniques:COLT\cite{colt} and Cluster\cite{cluster}. To adapt to diverse memory mappings, Anchor\cite{anchortlb} dynamically change the density of anchor entries to exploit the contiguity. Based on the observation that TLB coalescing requires OS to allocate the certain sizes of contiguity chunks, Ranger\cite{rangar} provides an OS service, which can allocate large contiguity chunks even with memory fragmentation by coalescing scattered physical frames into contiguous regions. Direct segment \cite{efficient} adds a variable-sized HW segment to represent a large range of contiguous pages. RMM\cite{RMM} supports multiple segments by adding a redundant TLB and corresponding page table. DVM\cite{DVM} extends the concepts of direct segments for area-constrained accelerators.
For multicore systems, many approaches leverage the shared translation across multiple cores to reduce TLB misses by exploiting the sharing data\cite{bhattacharjee2011shared,li2011leveraging,srikantaiah2010synergistic}.
Recently, \cite{cox2017efficient} proposes Mix TLB that supports multiple page sizes with a single indexing scheme.

\para{Reducing TLB Miss Penalty and Prefetching.} The first primary approaches decrease the penalty of page table walk by improving the translation cache\cite{barr2010translation,bhattacharjee2013large}.
which reduces the number of memory accesses to fetch intermediate page table nodes.
\cite{barr2011spectlb} interpolates speculated address translation based on the local pages using a reservation- based physical memory allocator. As the predictable behaviors of memory accesses, prefetching is also be used on address translation by proactively inserting pages used in the near future \cite{navarro2002practical,prefetch2,saulsbury2000recency}. 
Virtual caching is another direction to reduce address translation overheads and it allows to delay address translation after cache misses. Many existing works deployed to virtual caching can reduce page table walks as large on-chip caches can contain data which could have missed in TLBs of the conventional physical caching \cite{basu2012reducing,Efficientsynonym,yoon2016revisiting,zhang2010enigma}.

%\para{Virtualized Systems.}
%As virtual machines add an additional layer of address translation, TLB miss latencies are amplified, and thus the virtualized systems exhibit more severe performance drops by TLB misses, compared to native ones. Various prior work have tackled the translation challenge of virtualization. Gandhi et al. extended the segment-based translation to support nested translation of virtualized systems\cite{gandhi2014efficient}. 
%On the other axis, there were studies that improve the TLB miss latency handling by improving the translation cache\cite{bhargava2008accelerating} or improving the organization of the nested page tables \cite{ahn2012revisiting,gandhi2016agile}.

\section{Conclusion}

The introduction of mixed contiguity brings challenge to existing TLB coverage improvement techniques. To mitigate this issue, we propose $\mathbf{K}$ Aligned TLB, a SW-HW hybrid TLB coalescing scheme, which exploits the complicated distribution of contiguity chunks of memory mappings using different sizes of container ($\mathbf{K}$-bit aligned entries) and maintains the translation efficiency by adding the algorithm of determining $\mathbf{K}$ and  speculation for TLB lookup. With diverse contiguities of memory mappings,  $\mathbf{K}$ Aligned  provides a better performance than existing related works.

\bibliographystyle{plain}
\bibliography{Ref}

\end{document}